\documentclass[prb,twocolumn,showpacs,superscriptaddress]{revtex4}

\usepackage{graphicx}
\usepackage{dcolumn}
\usepackage{amsmath}

\begin{document}

\begin{abstract}

We have used tight-binding molecular-dynamics simulations to investigate the
role of point defects (vacancies and interstitials) on structural relaxation
in amorphous silicon. Our calculations give unambiguous evidence that point
defects can be defined in the amorphous solid, showing up as anomalies in the
valence-charge/Vorono\"{\i}-volume relation. The changes in the radial
distribution functions that take place during annealing are shown to be in
close agreement with recent, highly-accurate x-ray diffraction measurements.
Our calculations provide strong evidence that structural relaxation in {\em
a}-Si proceeds by the mutual annihilation of vacancies and interstitials,
i.e., local structural changes rather than an overall relaxation of the
network.


\begin{center}
Submitted to {\em Physical Review Letters} \\
\end{center}

\end{abstract}


\title{\bf Point defects in models of amorphous silicon and their role in
structural relaxation}

\author{\firstname{Cristiano L.} \surname{Dias}} \email[E-mail:
]{cl_dias@yahoo.com} \affiliation{D\'{e}partement de Physique et
Groupe de Recherche en Physique et Technologie des Couches Minces
(GCM), Universit\'{e} de Montr\'{e}al, Case Postale 6128,
Succursale~Centre-Ville, Montr\'{e}al, Qu\'{e}bec, Canada H3C 3J7}
\author{\firstname{Laurent J.} \surname{Lewis}} \email[To whom
correspondence should be addressed; e-mail: ]
{Laurent.Lewis@UMontreal.CA} \affiliation{D\'{e}partement de Physique
et Groupe de Recherche en Physique et Technologie des Couches Minces
(GCM), Universit\'{e} de Montr\'{e}al, Case Postale 6128,
Succursale~Centre-Ville, Montr\'{e}al, Qu\'{e}bec, Canada H3C 3J7}
\author{\firstname{S.} \surname{Roorda}} \email[E-mail:
]{sjoerd.roorda@umontreal.ca} \affiliation{D\'{e}partement de Physique
et Groupe de Recherche en Physique et Technologie des Couches Minces
(GCM), Universit\'{e} de Montr\'{e}al, Case Postale 6128,
Succursale~Centre-Ville, Montr\'{e}al, Qu\'{e}bec, Canada H3C 3J7}

\date{\today}

\maketitle

\narrowtext

The amorphous phase of tetrahedral covalent semiconductors is usually
described in terms of an ideal continuous random network (CRN),\cite{Zallen}
which is fully connected and perfectly fourfold coordinated; in this model,
the angular and radial distributions differ only slightly from their
crystalline counterparts at short range. Yet, the atomic structure of these
materials --- and in particular that of the prototypical amorphous silicon
({\em a-}Si) --- has not been completely resolved and the analogy with the
CRN not fully established. In particular, the presence of point defects in
the network and their role during structural relaxation remains unclear. This
is the problem we address here.

A striking similarity has been noted in {\em a-}Si between the changes
induced by annealing (structural relaxation) and those associated with the
removal of damage caused by radiation in {\em crystalline} Si.\cite{Roorda}
This has led to the suggestion that point defects such as vacancies and
interstitials (i) do exist in {\em a-}Si and (ii) are closely related to
structural relaxation. Experimental evidence for the existence of such
defects, and their role in relaxation, is based on the kinetics of heat
release upon annealing,\cite{Roorda} M\"ossbauer spectroscopy,\cite{Mossbauer}
Cu solubility and diffusion,\cite{Diffusion} and, more recently, high-energy
X-ray diffraction (XRD) measurements.\cite{Laaziri} In the latter experiment,
the radial distribution function (RDF) of {\it pure}, ion-implanted {\em
a-}Si was measured with very high precision, both before and after annealing.
Both samples were found to be under-coordinated and {\em equally dense}; the
average coordination number was found to increase from 3.79 to 3.88 upon
annealing. This, as well as other details of the RDFs, was interpreted in
terms of the mutual annihilation of vacancies and interstitials during
relaxation, which brings about an increase in the average coordination number
without affecting the density.

The concept of point defects provides an attractive perspective for
understanding the structure of amorphous semiconductors, since it is an
extension of well-known ideas developed for the corresponding crystalline
phases. However, some questions remain before this interpretation can be
fully accepted. In particular, can theory provide further evidence for the
existence of point defects in these materials, and can they be held
responsible for structural relaxation? What are the structural and electronic
properties of vacancies and interstitials in amorphous semiconductors? In
this Letter, we demonstrate, based on quantum-mechanical tight-binding
simulations, that the interpretation of structural relaxation in {\em a-}Si
in terms of point defects is indeed correct. We propose specific criteria for
identifying point defects and show that the changes in the RDF observed in
the XRD experiments can be reproduced in detail if defects are assumed to
exist and to annihilate.

Our study is based on an atomistic description of the material. We employed,
as a starting point, a CRN model containing 216 atoms prepared using a
modified Wooten-Winer-Weaire algorithm\cite{Mousseau} and the empirical
Stillinger-Weber potential. This sample was then relaxed using tight-binding
molecular dynamics\cite{TBMD} (TBMD) at constant volume. The TBMD scheme
ensures a proper, and ``affordable'', description of the energetics of the
material, while also providing information on the electronic structure; {\em
ab initio} simulations are out of reach given the size of the simulations
needed for the present study. The specific implementation of TBMD we used is
that of Kwon {\em et al},\cite{Kwon} which is cut-off at $4.18$ \AA, i.e.,
beyond the second-neighbour shell. The TBMD-relaxed sample has a density of
0.047 atoms \AA$^{-3}$, is perfectly fourfold coordinated, and the width of
the angular distribution is a minuscule 10.30 $^\circ$. This model will be
taken as representing the fully-relaxed material.

In order to assess the role of point defects on relaxation, we also prepared
``as-made'' samples by introducing, by hand, a number of defects in the
fully-relaxed sample. A vacancy is created, simply, by removing one atom from
its equilibrium position, and an intestitial by adding an atom to the CRN.
For the latter case, there exist many possibilities; we chose configurations
closest to the dumbell, as this is the one with the lowest energy in the
crystal.\cite{Colombo} After the introduction of each defect, the system was
fully relaxed. For vacancies, a static relaxation was found to be sufficient.
For interstitials, the relaxation is sensitive to initial conditions; a
simulated annealing algorithm was used to scan the energy landscape for a
global minimum by quenching slowly from 300 K to 0 --- the temperature was
scaled down by a factor of 0.997 at every timestep $= 1.08 \times 10^{-15}$
s. It should be stressed that a detailed study of the complete path from the
as-made to the fully-relaxed material is beyond the reach of atomistic
simulations.

Altogether, 40 samples, each containing a single vacancy or a single
interstitial (located at random) were prepared. In order to demonstrate that
defects do exist, it is necessary to be able to locate them with reasonable
confidence in the structure, {\em without a priori knowledge}. Evidently, no
criterion will give a perfect score, as there will always be marginal cases
due to the continuously disordered nature of the material. We have found that
a criterion based solely on distance --- e.g., to identify clusters of
threefold atoms which may signify the presence of a vacancy --- are not
sufficiently robust for this purpose. Nevertheless, it is possible to
identify defects by looking for correlations between local properties; the
assumption, which we have verified, is that the presence of a defect is felt
more strongly at short range. In particular, defects are expected to have a
sizeable effect on the volume available to neighbouring atoms\cite{Raymond}
--- the Vorono\"{\i} volume --- as well as on their valence
charges.\cite{Kim1,Kim2}

To illustrate this point, we plot, in Figs.\ 1(a) and (b), the Vorono\"{\i}
volumes and valence charges of all atoms for the fully-relaxed sample
(crosses) as well as two defective samples, each containing a single vacancy
(squares); as will be shown below, these particular samples are typical of
the set we generated. {\em Modulo} some dispersion, a clear correlation can
be observed, {\em for most atoms}, between the volume and the valence charge.
Both quantities are well bounded and the valence charge decreases with the
Vorono\"{\i} volume. This can be understood as follows: The Vorono\"{\i} volume is
proportional to the cube of the mean nearest-neighbour distance; thus, small
Vorono\"{\i} volumes are related to large ion-ion repulsions which can only lead
to stable structures if screened by large amounts of valence electrons.

{\em For some atoms}, however, the correlation fails: some points --- all of
which belonging to the defective samples --- clearly fall outside the main
region. Examination of ball-and-stick models of the structures (insets)
reveals that these atoms are spatially correlated, and in fact sit near the
sites where the vacancies were created. We note that in one case, Fig.\ 1(b),
the ``peculiar'' atoms do not possess a particularly large volumes (two of
them are fivefold coordinated and the other is fourfold), while they do in
the case of Fig.\ 1(a) (with all four atoms threefold coordinated); the
Vorono\"{\i} volume, therefore, is {\em not} a good probe of the presence of
vacancies, in contrast to the situation in crystalline material.

Having established that vacancies can be identified in {\em a-}Si without
{\em a priori} knowledge of their position, we now present a statistical
analysis of all 20 samples containing a vacancy. Fig.\ 1(c) shows the
correlation between valence charge, Vorono\"{\i} volume, and coordination (as
determined by a distance cutoff argument). Again, most atoms are found to
belong to the main region, but a number of data points fall outside of it.
Ball-and-stick models of the atomic structures reveal, again, that these
points correspond to atoms surrounding vacancies. The particular examples of
Fig. 1(a) and 1(b) are thus representative of all samples studied. Similar
observations apply to interstitial-type defects, which we do not show here
for lack of space.

The distribution of points outside the main region is found to correlate to
coordination: Threefold-coordinated atoms (squares) tend to have large
volumes and exhibit a ``bimodal'' distribution of valence charge --- either
large or small compared to the atoms in the main region. Fivefold atoms
(circles), in contrast, occupy relatively small volumes and are surrounded by
low amounts of valence charge. The volume and valence charge of
fourfold-coordinated atoms, finally, present no peculiarity; their
distribution must be taken as inherent to the amorphous structure. Thus,
while vacancies are predominantly associated to threefold-coordinated atoms,
as is the case in crystalline systems, fivefold-coordinated atoms can also
signify the presence of a vacancy in the amorphous phase.

We have demonstrated that defects are stable entities that can be identified
in amorphous structures. We move on to assess their role in relaxation by
comparing the RDFs of models for the as-implanted and the annealed states,
displayed in Fig.\ 2, and constructed as follows: We first define a {\em
local}, atom-dependent RDF, $J_i(r)$, which is the number of particles per
unit length at distance $r$ from atom $i$. This can then be used to compute
the RDF associated to a particular vacancy, which is done by summing the
$J_i(r)$ of all atoms located within a distance $d$ from the vacancy; this
will be labeled $J^d_{\rm V}(r)$. Likewise, we may define a $J^d_{\rm I}(r)$
for interstitials. We may now combine $J^d_{\rm V}$ and $J^d_{\rm I}$ into
the {\em total} RDF $J(r)$ of a sample containing $N_{\rm V}$ vacancies and
$N_{\rm I}$ interstitials, on average $2d$ apart:
   \begin{equation}
   J(r)= n_{\rm V} J^d_{\rm V}(r) + n_{\rm I} J^d_{\rm I}(r)
   \label{eq1}
   \end{equation}
where $n_{\rm V} = N_{\rm V}/(N_{\rm V}+N_{\rm I})$ and $n_{\rm I} = N_{\rm
I}/(N_{\rm V}+N_{\rm I})$. The latter expression is formally equivalent to
the usual RDF for a sample containing a density $\rho$ of defects with
relative concentrations of $n_{\rm V}$ vacancies and $n_{\rm I}$
interstitials. For proper statistics, $J^d_{\rm V}$ and $J^d_{\rm I}$ were
averaged over many samples --- 20 with a vacancy and 20 with an interstitial.
There is a single defect per sample and defect-defect interactions are
ignored. Our results are therefore valid in the limit of low defect
densities, the latter being determined by the value of $d$ as
$\rho=1/\frac{4}{3} \pi d^3$.

We now examine the structural changes brought about by the annihilation of
defects. We assume here that the structure of as-implanted and annealed
samples are similar, except for the defect density and relative
concentration. Thus, we can mimic both states of the material by simply
varying $d$ and $n_{\rm V}$ (or $n_{\rm I}$). The precise form of the
evolution from one state to the other can only be obtained through a detailed
analysis of the atomic mechanisms that take place during annealing, which is
beyond the reach of current simulations.

As noted earlier, the XRD measurements of Laaziri {\em et al}.\cite{Laaziri}
suggest that the structural relaxation between as-implanted and annealed
samples proceeds by the annihilation of point defects, more precisely the
diffusion and recombination of vacancies and interstitials, which not only
cause the coordination to increase, but also is such that the density is
conserved. Thus, it must be the case that
   \begin{equation}
   N^{\rm ann}_{\rm V} = N^{\rm imp}_{\rm V}-N^{\rm imp}_{\rm I},
   \label{eq5}
   \end{equation}
where $N^{\rm imp}_{\rm V}$ and $N^{\rm imp}_{\rm I}$ are the number of
vacancies and interstitials in the as-implanted sample, respectively, and
$N^{\rm ann}_{\rm V}$ is the number of vacancies in the annealed material; we
assume here that all interstitials annihilate with vacancies, $N^{\rm
ann}_{\rm I}=0$, which imposes $n^{\rm imp}_{\rm V} \geq 0.5$.

Because density, and thus volume, is conserved during annealing, it must also
be the case that
   \begin{equation}
   N^{\rm ann}_{\rm V} (d^{\rm ann})^3 = (N^{\rm imp}_{\rm V} +
                                          N^{\rm imp}_{\rm I}) (d^{\rm imp})^3
   \label{eq2}
   \end{equation}
where $2d^{\rm imp}$ and $2d^{\rm ann}$ are the average distances between
defects in the as-implanted and annealed sample, respectively. Combining
eqs.\ \ref{eq2} and \ref{eq5}, we thus have
   \begin{equation}
   d^{\rm ann} = d^{\rm imp} \sqrt[3] {\frac{N^{\rm imp}_{\rm V}+N^{\rm imp}_{\rm I}}
                                            {N^{\rm imp}_{\rm V}-N^{\rm imp}_{\rm I}}}
   \label{eq3}
   \end{equation}
The above constraints reduce from 6 to 3 the number of parameters needed to
calculate the RDF of the as-implanted and annealed materials, viz.\ $N^{\rm
imp}_{\rm V}$, $N^{\rm imp}_{\rm I}$ and $d^{\rm imp}$.

We now demonstrate that the mutual annihilation of vacancies and
interstitials are responsible for the changes observed during structural
relaxation, i.e., increase in coordination number of the first and second
neighbour peaks in the RDF and corresponding decrease of the ``noise''
in-between those peaks. We show in Fig.\ 2 the simulated RDFs for the
as-implanted and annealed models, using $n^{\rm imp}_{\rm V}=\frac{2}{3}$
(and thus $n^{\rm imp}_{\rm I}=\frac{1}{3}$) and $d^{\rm imp}=5.5$ \AA. The
corresponding densities of defects are 3.1 at.\% for the as-implanted
material and 1.0 at.\% for the annealed one, of the same order as those
inferred from the experiments of Laaziri {\em et al.}\cite{Laaziri}

We indeed observe, in Fig.\ 2, a decrease of the ``noise'' in between the
first and second-neighbour peaks and an increase in the amplitude of the
second peak following the annihilation of defects. Quantitatively, the
coordination of the second peak --- computed by fitting the first half to a
Gaussian --- increases by 0.43, in good agreement with experiment (0.28)
considering the ``error bars'' inherent to both approaches.

Concerning the first peak, now, an analysis of the as-implanted sample shows
that the large-$r$ tail is overestimated when compared to experiment. This is
to a large extent an artifact of the computational approach, as complete
relaxation of the defects --- in particular the interstitials --- is
extremely difficult. In order to eliminate these artifacts, we follow the
same procedure as that used to analyze the experimental RDF: we evaluate the
coordination of the nearest-neighbour shell by fitting the first peak to a
Gaussian lineshape and integrating. We find the coordination to increase by
0.06 from $Z=3.91$ in the as-implanted sample to $Z=3.97$ in the annealed
sample. This difference is very significant: for $Z=3.91$, one atom in 11 is,
on average, threefold coordinated, while the proportion drops to one in 33
for $Z=3.97$. Taking error bars into account, this change in coordination is
in excellent agreement with the experimental value of $3.88-3.79 = 0.09$. It
should be noted that the precise correspondance between experimental and
model as-implanted samples is impossible to establish; we are really only
interested in relative changes upon annealing.

The agreement between model and experiment --- and thus the interpretation of
relaxation in terms of point-defect annihilation --- is unambiguously
established in the inset of Fig.\ 2, where we plot the {\em difference}
between annealed and as-implanted material for both experiment (full line)
and the present model (dotted line). In order to set a common reference, the
positions of the first peak of the RDFs were forced to match. The coincidence
between the two data sets is striking, even though the amplitudes of the
various peaks is not perfect, due to the approximate character of the
computational approach. The reduction of noise between first- and
second-neighbour peaks is qualitatively reproduced, and even such fine
details as the small peaks or shoulders in the second-neighbour shell
(3.25--4.0 \AA), indicated by arrows, are correctly reproduced. These results
provide strong evidence that point-defect annihilation is, indeed, the
mechanism responsible for structural relaxation in amorphous silicon.

In summary, we have demonstrated, based on TBMD calculations of point defects
in amorphous silicon, that there exists a correlation between the valence
charge and the Vorono\"{\i} volume that is broken in the presence of defects.
Comparison between model as-implanted and annealed samples provides strong
evidence for the interpretation of structural relaxation in {\em a}-Si in
terms of the mutual annihilation of vacancies and interstitials, which
recombine such as to increase the nearest-neighbour coordination while
keeping the density constant. Thus, the concept of point defects is relevant
to amorphous silicon. Annealing evidently proceeds by local structural
changes rather than an overall relaxation of the network.

{\it Acknowledgments} -- It is a pleasure to thank Normand Mousseau and Ralf
Meyer for useful discussions. This work is supported by grants from the
Natural Sciences and Engineering Research Council (NSERC) of Canada and the
``Fonds pour la formation de chercheurs et l'aide {\`a} la recherche'' (FCAR)
of the Province of Qu{\'e}bec. We are indebted to the ``R\'eseau qu\'eb\'ecois de
calcul de haute performance'' (RQCHP) for generous allocations of computer
resources.


\begin{figure}[t]
\includegraphics*[width=10cm]{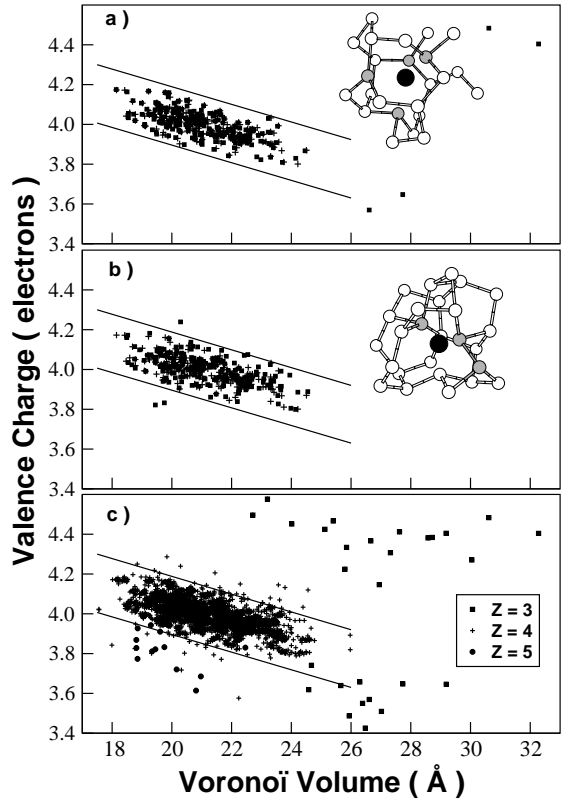}
\caption{
Valence charge vs Vorono\"{\i} volume for samples with and without vacancies. (a)
and (b) are for two particular cases. Crosses are for defect-free samples
while squares are for samples containing a single vacancy. The ball-and-stick
models show the spatial correlation which exists between the position of the
vacancy (black circle) and the positions of the atoms exhibiting an anomalous
charge-volume relation (grey circles). In (c), the corresponding data is
collected for all 20 samples containing a vacancy, and are sorted according
to coordination, as indicated. The full lines --- which are the same in all
three panels --- are a guide to the eye delimiting the ``normal'' region of
correlation.
}
\label{fig1}
\end{figure}

\begin{figure}[t]
\vspace*{-5cm}
\includegraphics*[width=10cm]{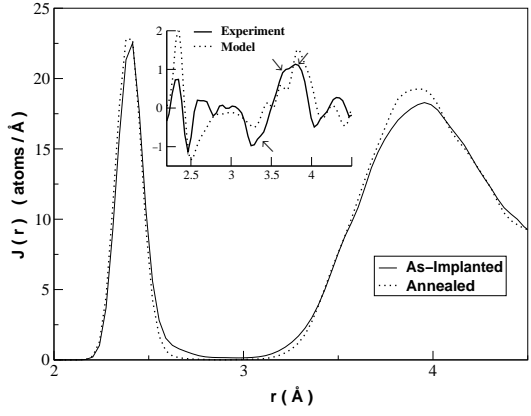}
\caption{
RDFs of as-implanted and annealed models, for a relative concentration of
vacancies of $n_{\rm V}^{\rm imp}=\frac{2}{3}$ and $d^{\rm imp}=5.5$ \AA. The
inset shows the {\em difference} between annealed and as-implanted RDFs for
both model and experiment.
}
\label{fig2}
\end{figure}

\end{document}